\input harvmac.tex
\def\Title#1#2{\rightline{#1}
\ifx\answ\bigans\nopagenumbers\pageno0\vskip1in%
\baselineskip 15pt plus 1pt minus 1pt
\else
\def\listrefs{\footatend\vskip 1in\immediate\closeout\rfile\writestoppt
\baselineskip=14pt\centerline{{\bf References}}\bigskip{\frenchspacing%
\parindent=20pt\escapechar=` \input
refs.tmp\vfill\eject}\nonfrenchspacing}
\pageno1\vskip.8in\fi \centerline{\titlefont #2}\vskip .5in}
%
%
\def\tx{\tilde x}\def\x{x}
\def\ty{\tilde\psi}\def\y{\psi}%
\def\Re{{\rm Re\ }}
\def\Im{{\rm Im\ }}
\def\I{I}
\def\II{\relax{\rm I\kern-.1em I}}
\def\IIa{{\II}A}
\def\IIb{{\II}B}
\def\CM{{\cal M}}
\def\CN{{\cal N}}
\def\eps{\varepsilon}
\def\ket#1{|#1\rangle}
\def\IR{\relax{\rm I\kern-.18em R}}
\def\BR{\IR}
\def\p{\partial}
%
\font\zfont = cmss10 
\font\zfonteight = cmss8 
\def\ZZ{\hbox{\zfont Z\kern-.4emZ}}
\def\sZZ{\hbox{\zfonteight Z\kern-.4emZ}}
\def\RR{\hbox{\zfont  l\kern-.1emR}}

\Title{\vbox{\baselineskip12pt
\hfill{\vbox{
\hbox{RU-96-44\hfil}
\hbox{hep-th/9606139}}}}}
{\vbox{\centerline{Branes Intersecting at Angles}}}
\centerline{Micha Berkooz, Michael R. Douglas and Robert G. Leigh}
\smallskip
{\it
\centerline{Department of Physics and Astronomy}
\centerline{Rutgers University, Piscataway, NJ 08855-0849}
}
\bigskip\centerline{\tt berkooz, mrd, leigh@physics.rutgers.edu}
\bigskip\bigskip
\baselineskip 18pt
We show that configurations of multiple D-branes related by $SU(N)$
rotations will preserve unbroken supersymmetry.
This includes cases in which two D-branes are related by a rotation
of arbitrarily small angle, and we discuss some of the physics of this. In
particular, we discuss a way of obtaining 4D chiral fermions on the 
intersection of D-branes.

We also rephrase the condition for unbroken supersymmety as the
condition that a `generalized holonomy group' associated with the
brane configuration and manifold is reduced, and relate this condition
(in Type \IIa\ string theory) to a condition in eleven dimensions.


\Date{June 1996 (revised)}
%
\def\npb#1#2#3{Nucl. Phys. {\bf B#1} (#2) #3}
\def\plb#1#2#3{Phys. Lett. {\bf #1B} (#2) #3}
\def\prd#1#2#3{Phys. Rev. {\bf D#1} (#2) #3}
\def\prl#1#2#3{Phys. Rev. Lett. {\bf #1} (#2) #3}
\def\mpl#1#2#3{Mod. Phys. Lett. {\bf A#1} (#2) #3}
\def\hepth#1{hep-th/#1}
\def\nl{\noindent}

\lref\ghm{M. Green, J.A. Harvey and G. Moore, ``{\it I-Brane Inflow and 
Anomalous Couplings on D-Branes,}'' \hepth{9605033}.}

\lref\inflow{J. Blum and  J.A. Harvey, 
``{\it Anomaly Inflow for Gauge Defects,}''
\npb{416}{1994}{119}, \hepth{9310035};
\nl  J.M. Izquierdo and P.K. Townsend,
``{\it Axionic Defect Anomalies and their Cancellation,}''
\npb{414}{1994}{93}, \hepth{9307050}.}

\lref\sixd{E.G. Gimon and J. Polchinski, 
``{\it Consistency Conditions for Orientifolds and D-Manifolds,}'' 
\hepth{9601038};
\nl M. Berkooz, R.G. Leigh, J. Polchinski, J.H. Schwarz, N. Seiberg 
and E. Witten, 
``{\it Anomalies, Dualities and Topology of $D=6$ $N=1$ Superstring Vacua,}'' 
RU-96-16, \hepth{9605184};
\nl E.G. Gimon and C.V. Johnson, 
``{\it K3 Orientifolds,}'' 
NSF-ITP-96-16, hep-th/9604129; 
\nl A. Dabholkar and J. Park, ``{\it  Strings on Orientifolds,}'' 
CALT-68-2051, \hepth{9604178}; 
\nl A. Dabholkar and J. Park, 
``{\it An Orientifold of Type IIB Theory on K3,}'' 
CALT-68-2038, \hepth{9602030}.}

\lref\duality{
J.~H.~Schwarz and A.~Sen, 
``{\it Duality Symmetries of 4D Heterotic Strings,}''
\plb{312}{1993}{105}, \hepth{9305185};
\nl M.~J.~Duff, and R.~Khuri,
``{\it Four-Dimensional String-String Duality,}''
\npb{411}{1994}{473}, \hepth{9305142};
\nl C. Hull and P. Townsend, 
``{\it Unity of Superstring Dualities,}''
\npb{438}{1995}{109}, \hepth{9410167};
\nl A. Strominger, 
``{\it Massless Black Holes and Conifolds in String Theory,}''
\npb{451}{1995}{96}, \hepth{9504090}.}

\lref\schw{J. Schwarz, 
``{\it An $SL(2,Z)$ Multiplet of Type IIB Superstrings,}''
\plb{360}{1995}{13}, \plb{364}{1995}{252}, \hepth{9508143}.}

\lref\mtheory{E. Witten, 
``{\it String Theory Dynamics In Various Dimensions,}''
\npb{443}{1995}{85}, \hepth{9503124};
\nl P.K. Townsend,
``{\it The Eleven-Dimensional Supermembrane Revisited,}''
\plb{350}{1995}{184}, \hepth{9501068}.}

\lref\mumford{D. Mumford, {\it Tata Lectures on Theta, Vol. 1},
Birkhauser, 1983, pp. 44-66.}

\lref\dbrane{J. Dai, R.G. Leigh and J. Polchinski,
``{\it New Connections between String Theories,}'' \mpl{4}{1989}{2073}.}

\lref\dbi{R.G. Leigh, 
``{\it Dirac-Born-Infeld Action from Dirichlet $\sigma$-model,}''
\mpl{4}{1989}{2767}.}

\lref\joe{J. Polchinski, 
``{\it Dirichlet-Branes and Ramond-Ramond Charges,}''
\prl{75}{1995}{4724}, \hepth{9510017}.}

\lref\bsv{M. Bershadsky, V. Sadov and C. Vafa,
``{\it  D-Strings on D-Manifolds,}''
\npb{463}{1996}{398}, \hepth{9510225}.}

\lref\pcj{J. Polchinski, S. Chaudhuri, and C. V. Johnson, 
``{\it Notes on D-Branes,}'' NSF-ITP-96-03, \hepth{9602052}.}

\lref\andy{A. Strominger, ``{\it Open P-Branes,}''  \hepth{9512059}.}

\lref\miao{M. Li, ``{\it Boundary States of D-Branes and Dy-Strings,}''
\npb{460}{1996}{351}, \hepth{9510161}.}

\lref\mike{M. Douglas, ``{\it Branes Within Branes,}'' 
RU-95-92, \hepth{9512077}.}

\lref\bbs{K. Becker, M. Becker and A. Strominger,
``{\it Fivebranes, Membranes and Non-Perturbative String Theory,}''
\npb{456}{1995}{130}, \hepth{9507158}.}

\lref\cjs{E.~J.~Cremmer, B.~Julia and J.~Scherk, 
``{\it Supergravity Theory in 11 Dimensions,}'' \plb{5}{1978}{409}.}

\lref\bvs{M. Bershadsky,  V. Sadov and C. Vafa, 
``{\it D-branes and Topological Field Theories,}''  
\npb{463}{1996}{420}, \hepth{9511222}.}

\lref\bps{E. B. Bogomolny, ``{\it Stability of Classical Solutions,}''
Sov. J. Nucl. Phys. {\bf 24} (1976) 449;
\nl M. K. Prasad and C. M. Sommerfield, ``{\it An Exact Classical 
Solution for the 't Hooft Monopole and the Julia-Zee Dyon,}'' 
\prl{35}{1975}{760}.}

\lref\blfour{M. Berkooz and R.G. Leigh, 
``{\it A D=4 N=1 Orbifold of Type I Strings,}'' 
RU-96-28, \hepth{9605049}.}

\lref\interd{A. Sen, ``{\it U-duality and Intersecting D-branes,}'' 
\prd{53}{1996}{2874}, \hepth{9511026}.}

\lref\green{M. B. Green, C. Hull and P. K. Townsend,
``{\it D-brane Wess--Zumino actions, T-duality and the 
cosmological constant,}'' \hepth{9604119}.}

\lref\town{P. K. Townsend, \plb{350}{1995}{184},
\hepth{9501068}.}

\lref\interm{G. Papadopoulos and P. K. Townsend,
``{\it Intersecting M-branes,}'' 
DAMTP-R-96-12, \hepth{9603087};
\nl J.P. Gauntlett, D.A. Kastor and J. Traschen,
``{\it Overlapping Branes in M Theory,}''
CALT-68-2055, hep-th/9604179;
\nl N. Khviengia, Z. Khviengia, H. Lu and C. N. Pope
``{\it Intersecting M-branes and Bound States,}''
CTP-TAMU-19-96, \hepth{9605077}.}

\lref\ooguri{H. Ooguri, Y. Oz and Z. Yin,
``{\it D-Branes on Calabi-Yau Spaces and Their Mirrors},''
hep-th/9606112.}

\lref\duff{M. J. Duff, B. E. W. Nilsson and C. N. Pope, 
Phys. Rep. {\bf 130} (1986) 1, and references therein.}

\lref\nicolai{
B. de Wit and H. Nicolai, \npb{274}{1986}{363};
\nl H. Nicolai, \plb{187}{1987}{316}. }

\lref\duffstelle{M. J. Duff and K. Stelle, \plb{253}{1991}{113}.}

\lref\vbrgl{V. Balasubramanian and R.G. Leigh, 
``{\it More Branes at Angles},'' to appear.}

\newsec{Introduction}

Dirichlet (D) branes are BPS states and
break half of the available supersymmetries \refs{\joe,\pcj}.
By adding D-branes filling space-time to a string compactification,
one thereby obtains models with additional gauge symmetry and
less supersymmetry.
Configurations containing multiple D-branes break supersymmetry 
further, and can leave interesting fractions such as $1/4$ or $1/8$.
Such models have been studied in many recent works.
\refs{\bsv,\interd,\interm,\mike,\blfour,\sixd}

So far, only D-branes intersecting at right angles have been studied.
An indication that this is not the only possibility comes from 
considering the compactification on
the torus $T^4\times \BR^6$ with a $p$-brane
at a point in $T^4$ and a wrapped 
$(p+4)$-brane, a configuration which preserves $1/4$
of the supersymmetry.
One can T-dual any of the dimensions of the torus to produce
intersecting D-branes wrapped around some of its dimensions, for example 
$(p+2)$-branes wrapped around two $2$-planes.
If the metric components  $g_{ij}\ne 0$ for $i\ne j$, 
they will intersect at an angle. It is clear however that this is
still a BPS state.

In this paper, we study more general configurations of D-branes.
We find that the condition for intersecting D-branes to preserve
supersymmetry is closely related to the well-known condition for a
$d$-dimensional
curved manifold to preserve supersymmetry: a generalized holonomy
(which we will define) must be contained in a $SU(d/2)$ subgroup of 
$SO(d)$.
This condition can be satisfied with arbitrary angles of intersection.
We derive this condition and give examples of solutions in section 2.

Open strings joining two such D-branes are quasi-localized: as the
opening angle $\theta$ decreases, their allowed spread from the
intersection point increases.
In section 3, we consider the world-sheet analysis of these
strings, while in section 4 we discuss the possibility of describing
these configurations by starting with the space-time effective Lagrangian
for parallel D-branes and turning on a linearly increasing displacement.
We show that these quasi-localized open string states form a
Kaluza-Klein-like tower with spacing proportional to the angle.
Section 3 also discusses 4D chiral fermions on the intersection of
D-branes.

We then study T-duality,
and find that the angles between branes are dual to orthogonal
branes in certain spacetime backgrounds.
In the final section, we give an 11-dimensional interpretation of our
results.

\newsec{Supersymmetry}

Type \II\ string theory has $\CN=2$ supersymmetry in $d=10$ with
parameters $\eps$ and $\tilde\eps$, for left and right movers.
In the presence of a D-brane, an unbroken supersymmetry must 
satisfy\foot{The metric is $(-,+,+,....)$. $\Gamma_0=\bigl(\matrix{0&-1\cr 
+1&0}\bigr)$,
$\Gamma_i=\bigl(\matrix{0&\gamma_i\cr\gamma_i&0}\bigr)$, where $\gamma_i$ 
are
real symmetric matrices.}
\eqn\unb{\tilde\eps = \prod_i e^\mu_i \Gamma_\mu\ \eps}
where $e_i$ is an orthonormal frame spanning the D-brane.
Let us refer to this product of $\Gamma$-matrices as $\Gamma_D$.

In the presence of several D-branes, an unbroken supersymmetry must
satisfy the condition \unb\ for each brane.  
Two D-branes related by a rotation $R$ therefore lead to the condition
\eqn\pairunb{\Gamma_D\;\eps = R^{-1}\Gamma_D R\;\eps}
which will have a solution if 
\eqn\pairdet{\det(\Gamma_D-R^{-1}\Gamma_D R)=0.}

Clearly the common dimensions of the branes do not affect the argument,
so we will assume for definiteness that the time dimension is common
to all the branes. Hence, in what follows, when we refer to the
dimension
of a brane, we will give only the (Euclidean) dimensions of 
interest.\foot{Thus a 2-brane is really a $(p+2)$-brane, having $p+1$
additional dimensions. 
We trust that this will confuse the reader.}

As is usual in supersymmetric compactification, the results will be
most simply phrased in terms of complex structure of the target space.
We work in $N$ complex dimensional space with coordinates $z^i$.
We define raising and lowering operators
\eqn\cliff{\eqalign{
a^\dagger_k &= \half(\Gamma_{2k-1} - i\Gamma_{2k})  \cr
a^k &= \half(\Gamma_{2k-1} + i\Gamma_{2k})
}}
acting on the $2^N$-dimensional sum of the two spinor representations.
Define the `vacuum' $\ket{0}$, satisfying $a^k\ket{0}=0$.
\def\Rdag#1#2{\left. R^\dagger\right._#1^{\;\; #2}\; }
An $SU(N)$ rotation $z^i\rightarrow R^i_j z^j$
acts on the Clifford algebra as
\eqn\surot{\eqalign{
a^k &\rightarrow R^k_{\;\;l}\; a^l \cr
a^\dagger_k &\rightarrow \Rdag{k}{l} a^\dagger_l .
}}

If the
branes are embedded in (say) a Calabi-Yau,
this division into creation
and annihilation operators is determined by the manifold.
If not, we can choose complex coordinates in whatever fashion we like. 
We will find below that two branes preserve a common supersymmetry
when they are related by an $SU(N)$ rotation, and in flat space
this will mean that
there exist complex coordinates such that the rotation is $SU(N)$.

Another way to think about this result is as follows.
In curved space, a
single D-brane will preserve half of the supersymmetries if it
is a supersymmetric subspace \refs{\bbs,\bvs},
a condition which can be stated using the geometric data
of complex structure, volume form and so on.
(See \ooguri\ for a conformal field theory generalization of this statement.)
The flat space limit of this condition is just eq. \unb.
In flat space, a single D-brane always preserves half of the 
supersymmetry, so any hyperplane can be a supersymmetric subspace.
Once we use it as such, we constrain the possible complex structures
available for other supersymmetric subspaces.

The condition bears a close resemblance to the condition that a curved
manifold preserve a supersymmetry -- that its holonomy live in a subgroup
of $SO(d)$ such as $SU(d/2)$.  If we could think of the matrix $\Gamma_D$
of \unb\ as a discrete holonomy acting on the supersymmetry parameter
$\eps$, this analogy would become precise.  We will discuss a precise 
form
of this in the last section.

We proceed to discuss some explicit solutions.

\subsec{Holomorphic Curves}

A $2$-brane is supersymmetrically embedded into a holomorphic curve,
which in flat space must be a plane $v_i W^i$ ($i=1,\ldots, N$).
Given $K$ such $2$-branes, an $SU(N)$ rotation takes any holomorphic 
plane into any other.
All of these planes preserve a common supersymmetry.

Explicitly, embed the first brane
into the complex $W^1$ plane.
It now has
\eqn\gamtwo{\eqalign{
\Gamma_{D1} &= i(a^\dagger_1 + a^1)(a^\dagger_1 - a^1) \cr
&= i(1 - 2 a^\dagger_1 a^1).
}}
An $SU(N)$ rotation can take this into a general complex plane, with
\eqn\gamtwopr{
\Gamma_{D,R} = i(1 - 2 v^*\cdot a^\dagger v\cdot a).
}
All of these preserve the state $\ket{0}$, as well as the state
$\prod_{k=1}^N a^\dagger_k  \ket{0}$ (since $|| v ||^2=1$).\footnote*{
We thank A. Morosov for independently
verifying that such a configuration will
preserve a supersymmetry.}

Without loss of generality, let us consider $N=2$. 
The supersymmetries left by the
first brane is one from the ${\bf 2}_L$ and one from the
${\bf 2}_R$ representations of $SU(2)_L\times SU(2)_R\sim SO(4)$. 
The ${SU(2)}$ above is one of these $SU(2)$'s. 
There is in fact another configuration that does not break
all the supersymmetry which is a diagonal rotation in both $SU(2)$'s; 
this
corresponds to two separate rotations of the plane parallel and
perpendicular to the first brane, but there is a set of complex
coordinates in which this is an $SU(2)$ rotation.

\subsec{$n$-branes in dimension $2n$}

In $2n$ dimensions, it is natural to embed
$n$ dimensions of the world-volume of the first brane
into $\Re Z^i$.
Again, other branes related to the first by $SU(n)$ rotation
will preserve a common supersymmetry.

Proof: The first D-brane then has
\eqn\gamone{\Gamma_{D1} = \prod_{k=1}^n \left(a^\dagger_k + a^k\right).}
The condition \pairunb\ then becomes 
\eqn\supairunb{
\prod_k \left(a^\dagger_k + a^k\right)\eps = 
\prod_k \left( \Rdag{k}{l} a^\dagger_l + R^k_{\;\;l} \;a^l\right)\eps .
}
Now $\eps=\ket{0}$ is a solution:
if $R$ is special unitary, both sides of \supairunb\ are equal
to $\prod_k a^\dagger_k \ket{0}$.

The same argument shows that $\prod_k a^\dagger_k \ket{0}$ is
another unbroken supersymmetry.  For $n$ even these spinors have
the same chirality and two chiral supersymmetries are preserved, 
while for $n$ odd they have opposite chirality and one supersymmetry
is preserved.

\subsec{Remarks}

The simplest example of the above solutions consists of two 2-branes
in four dimensions $X^{6,7,8,9}$. 
Begin with two 2-branes oriented along the
$X^{6,8}$-axes, and then rotate one of them by the angle
$\theta$ in the $X^6X^7$ plane and $-\theta$ in the $X^8X^9$ plane. 
We will often use
this case in the discussion which follows.

In terms of section 2.2, we define the complex coordinates 
$Z^1=X^6+iX^7$ and $Z^2=X^8+iX^9$. The branes are related by
$Z^1\rightarrow e^{i\theta} Z^1$ and
$Z^2\rightarrow e^{-i\theta} Z^2$.

Conversely, in the construction of section 2.1, the complex coordinates 
are
$W^1=X^6+iX^8$ and $W^2=X^7-iX^9$
(the orientation must be consistent, dictating the minus sign).
The rotation is then a real element of $SO(2)\subset SU(2)$.

It is clear that $1$-branes can not be oriented at non-trivial angles
in a supersymmetric fashion.
There are also many supersymmetric configurations consisting of
branes of differing dimensions. Many of these are T-dual to the
above examples.

Type \I\ theory also contains $9$-branes which require the condition
$\tilde\eps=\eps$.
This is compatible with the $2$-brane solutions but not the $N$-brane
solutions in general. 

We can make the analogous conjecture for
the exceptional reduced holonomy groups $G_2$ or $Spin(7)$ as well --
configurations of branes related by rotations in these groups will 
preserve $1/8$ or $1/16$ supersymmetry.

\newsec{World-sheet results}

Given two branes related by a rotation $R$, we work with 
complex coordinates $z^i$ in which $R$ is diagonal with eigenvalues
$\exp i\theta_i$, and define corresponding world-sheet fields $Z^i$.
Then the boundary conditions on stretched open strings are (in the 
notation of section 2.2)
\eqn\branebc{\eqalign{
&\Re {\p\over\p\sigma} Z^i|_{\sigma=0} = 0 \cr
&\Im Z^i|_{\sigma=0} = 0 \cr
&\Re e^{i\theta_i}{\p\over\p\sigma} Z^i|_{\sigma=\pi} = 0 \cr
&\Im e^{i\theta_i}Z^i|_{\sigma=\pi} = 0 .
}}
These shift both bosonic and fermionic mode expansions by $\theta_i/\pi$.

\subsec{$n=2$}

Let us study the example of section 2.3, and let
$\alpha\equiv{\theta_1/\pi}>0$.
When $\alpha=1/2$, the NS sector fermions become periodic,
and the space-time massless bosons are in the spinor of $SO(4)$, reduced
to a doublet by the GSO projection.
We can follow the spectrum from this point by continuously varying 
$\alpha$.  

The mode expansion for the (complex) bosonic field is
\eqn\modeexp{
Z(w,\bar w)=\sum_{m\in\sZZ} \left\{ \x_{-\alpha+m}e^{i(m-\alpha)w}
+\tx_{\alpha+m}e^{-i(m+\alpha)\bar w}\right\}
}
where $w=\sigma+\tau$, $\bar w=\sigma-\tau$ and $\x,\tx\in\RR$. 
The oscillators have non-zero commutators 
$$[\x_{-r},\tx_s]={1\over r} \delta_{r,s}.$$ In the NS sector,
the fermion modes are shifted by a further $1/2$.
The vacuum energy is found to be $\Delta E=-{1\over2}+\alpha$.

We can form states by raising the vacuum with the oscillators
$\x_{-\alpha-m}$, $\tx_{\alpha-1-m}$, $\y_{-\alpha-1/2-m}$,
and $\ty_{\alpha-1/2-m}$ for $m\geq 0$, where we are assuming
$0\geq\theta\geq\pi/2$. The GSO-like projection will lift
the vacuum as well as bosonic excitations of it. Thus we get
massless bosonic states:
\eqn\masslst{
\ty^i_{\alpha-1/2}|0\rangle \;\;\;\; (i=1,2)
}
which we expect to be localized at the crossing of the two branes.
This result may be found in an operator formalism as well; the GSO
projection is found by imposing mutual locality. For example, in the
NS sector, there are the vertex operators
\eqn\vab{ \eqalign{
\sigma_+^1\sigma_-^2e^{i(1-\alpha)H_1}e^{i\alpha
H_2}e^{-\phi}\cr
\sigma^1_+\sigma^2_-e^{-i\alpha
H_1}e^{-i(1-\alpha)H_2} e^{-\phi}
}}
in this sector. 
The operator-state mapping should be clear from eqs. \masslst\ and \vab.
The sector in which the open string between the 2 D-branes is oriented
in the other direction will have the same amount of bosonic and fermionic
spacetime fields. These will have the bosonic twist operators 
$\sigma_-^1\sigma_+^2$ in their vertex operators.

For small $\alpha$, there are towers of bosonic states
\eqn\NStower{\eqalign{
\ty^i_{\alpha-1/2}
\left(\x^{1}_{-\alpha}\right)^{n_1}
\left(\x^{2}_{-\alpha}\right)^{n_2}
&|0\rangle
;\;\;\;\; n_1+n_2=N\cr
\y^i_{-\alpha-1/2}
\left(\x^{1}_{-\alpha}\right)^{n_1}
\left(\x^{2}_{-\alpha}\right)^{n_2}
&|0\rangle;\;\;\;\; n_1+n_2=N-2\cr
\y^\mu_{-1/2} 
\left(\x^{1}_{-\alpha}\right)^{n_1}
\left(\x^{2}_{-\alpha}\right)^{n_2}
&|0\rangle
;\;\;\;\; n_1+n_2=N-1\cr
}}
each with energy $E_{N}=\alpha N$. The corresponding towers in the
Ramond sector are
\eqn\Rtower{\eqalign{
\left(\x^{1}_{-\alpha}\right)^{n_1}
\left(\x^{2}_{-\alpha}\right)^{n_2}
&|\pm\mp\rangle
;\;\;\;\; n_1+n_2=N\cr
\y^1_{-\alpha}\y^2_{-\alpha}\left(\x^{1}_{-\alpha}\right)^{n_1}
\left(\x^{2}_{-\alpha}\right)^{n_2}
&|\pm\mp\rangle
;\;\;\;\; n_1+n_2=N-2\cr
\y^i_{-\alpha}\left(\x^{1}_{-\alpha}\right)^{n_1}
\left(\x^{2}_{-\alpha}\right)^{n_2}
&|\pm\pm\rangle
;\;\;\;\; n_1+n_2=N-1
}}
At the massless level, we find a hypermultiplet in six 
dimensions.\foot{Here we count both orientations of the open strings.} 
This spectrum appears to be anomalous; the resolution of this problem
is discussed in \refs{\ghm,\inflow}.
At the $Nth$ mass level, there are  $2N$ massive 
hypermultiplets ($\equiv$ hyper$+$vector).

As $\theta\rightarrow 0$, this tower of states comes down, and fills
out eight-dimensional representations.
The DN strings are confined with a potential
$(\theta X)^2/\alpha'$ which is becoming more shallow. 
We will return to this in section 4.

\subsec{$n=3$}

Next, we consider the configuration of section 2.2 with $n=3$. This is 
particularly
interesting, because the two branes can intersect in 3+1 dimensions.
Let us discuss a pair of 6-branes in type \IIa\ intersecting
on a 4-plane; we will find that the open strings joining them produce
chiral 4D matter content on the intersection. 

This may seem counterintuitive.
One orientation of the open string has $U(1)\times
U(1)$ charge $(1,-1)$, while the other has charge $(-1,1)$. 
Which charge should we assign to the chiral field ? 
Equivalently, what correlates the world-sheet orientation with the
space-time chirality of the fermions?
The GSO projection, of course, but how does it see world-sheet orientation?

Given an oriented $4k+2$ dimensional space, we can define a `left-handed
rotation' as follows.  Reduce the rotation to normal form, a product
of $2k+1$ rotations in two-dimensional planes.  To each two-plane,
associate the two-form $dx\wedge dy$ where the sign is chosen to make the
rotation in the $(x,y)$ plane left handed.
A left-handed rotation is then one in which the wedge product of these forms
is positive.  The GSO projection will correlate the four-dimensional
chirality with the chirality of the rotation relating the two branes
and thus the orientation.

More explicitly,
take a rotation $R$ which in diagonal form is $R=diag\left(
e^{i\theta_1}, e^{i\theta_2}, e^{i\theta_3}\right)$, where 
$\sum_i\theta_i=0$.
The boundary conditions are as written in eqs. \branebc. The mode 
expansions
for the three $Z^i$ are as in \modeexp, each depending on the appropriate
$\alpha_i$. The ground state energy is 
$\Delta E=-1/2 + \alpha_1 + \alpha_2$, assuming $\alpha_{1,2}\geq 0$.
In the NS sector, we find a single massless spacetime boson
$\tilde\psi^3_{-1/2-\alpha_3}\ket{0}$ and a single spacetime massless
fermion.
There is a tower of states at energies $E=n_1\alpha_1+n_2\alpha_2$, which
can
be interpreted as arising from a harmonic oscillator potential $V=1/2\sum_{i}
\left( {\theta_iZ^i\over\pi}\right)^2$.

The vertex operators  for the open string that stretches
between the two 6-branes are:

Orientation 1:
\eqn\ora{V_b=\sigma_+e^{i(\alpha_1H_2+\alpha_2H_3+(1-\alpha_1-\alpha_2)H_4)},}
$$V_f=\sigma_-e^{i( (-{1\over2}+\alpha_1)H_2+(-{1\over2}+\alpha_2)H_3 +
({1\over2}-\alpha_1-\alpha_2)H_4)}e^{\pm i(H_0+H_1)}$$

Orientation 2:
\eqn\orb{V_b=\sigma_-e^{-i(\alpha_1H_2+\alpha_2H_3+(1-\alpha_1-\alpha_2)H_4)},}
$$V_f=\sigma_-e^{i( ({1\over2}-\alpha_1)H_2+({1\over2}-\alpha_2)H_3 +
(-{1\over2}+\alpha_1+\alpha_2)H_4)}e^{\pm i(H_0-H_1)}$$
where for future reference we have not yet gone to lightcone and thus have
two fermionic states in each sector.

We have obtained one physical fermion and one physical boson degrees
of freedom in each orientation of the string that stretches between 
the two 6-branes. These assemble themselves into an N=1 chiral
multiplet.
The GSO projection can be verified by requiring locality with 
respect to the supercharge.  This makes 
the space time fermion twist operator $e^{\pm{i\over2}(H_0+H_1)}$
in one orientation and $e^{\pm{i\over2}(H_0-H_1)}$ in the other
(we have not yet gone to the
lightcone to cut the number of states by half). The charge
assigned to the chiral field is that of the orientation that
has the left handed fermion.

The theory is anomalous, but as discussed in \ghm, this is
compensated by 6-brane bulk contributions. To show this, we just need
to reproduce eq. (2.3) of Ref. \ghm, and the rest is as explained there. 
We need to show that 
$I=ch_{(1,-1)}(F)\hat A(R)=2ch_{1}(F_1)ch_{-1}(F_2)\hat A(R)$ where
on the LHS we have the Chern classes of both $U(1)$ fields in the $(1,-1)$ 
represention and on the RHS the product of each the Chern classes in 
each $U(1)$ in the one dimensional representation 
with the appropriate charge. To show
this we expand the relevant part of $I$, which is the the 6-form part,
and just show the equality order by order. 

More generally, if we include $N_1$ ($N_2$) branes of each type, we find
$U(N_1)\times U(N_2)$ gauge theory. The chiral multiplet now transforms
in the $({\bf N_1, \bar N_2})$ representation. 

\newsec{Space-time Interpretation}

The parallel separation of two $Dp$-branes
can also be described as the Higgs mechanism in the $p+1$-dimensional 
field
theory that describes the low energy excitations of the two branes
when they are coincident. 
In this theory,
we can describe two branes intersecting at an angle by turning on
a linearly rising vev.
For example, if we let $X^6$ be a coordinate in the branes and $X^7$
perpendicular, then we should turn on the corresponding fields
$X^7 = X^6 \sigma_3\tan\theta$.

When the angle $\theta$ is small, this is clearly well-described by
the low-energy effective field theory.
We now show that such a vev will
truncate the spectrum as we expect and will localize the remaining
excitations around the intersection of the two D-branes. 

Even though we are rotating one of the branes in such a way that the
system is still BPS saturated,
there is no corresponding modulus in the $p+1$-dimensional
field theory. This is because there is
no boundary condition on the field theory at infinity that
parameterizes different angles between the branes, and therefore there
is no moduli space and no massless particle associated with it. 
One could compactify these directions on, say, a torus with the branes
wrapping on cycles; however, 
the angle between the branes is related to the shape of the torus, and
is not an independent parameter of the brane system.

Let us consider the example of section 2.1, adding four additional
dimensions to the branes to make them $5$-branes each with a $2$-plane
in $\BR^4$.
We will be interested in the field $W^2=X^7-iX^9$ which gives the
separation of the two D5-branes; this is a hermitian matrix which we
take to be $W^2=\phi \Sigma$ with
$\Sigma=\Bigl(\matrix{1&0\cr0&-1\cr}\Bigr)$.
Rotating one brane relative to the other will be given
(to lowest order in $\theta$) by 
$\phi={\theta}\Bigl(\matrix{Z_1\cr&-Z_1}\Bigr)$.

To check that we don't have to turn on more fields, we study the
unbroken supersymmetries of this configuration.
It is sufficient to consider the $D=10$ supersymmetry transformation,
reduced to 6-dimensions. The transformation law for the gaugino is
$$\delta\chi^a= F_{mn}\Gamma^{mn}\eta.$$ 
Now, the $\phi$ given above lifts to 10-dimensions as the
gauge field $A^{z_2}$; hence the non-zero components of the field
strength are $F_{z_1{\bar z}_2}=\partial_{z_1}A^{z_2}$ and
$F_{{\bar z}_1z_2}$. Thus, the remaining supersymmetries $\eta$ are such 
that they are
annihilated by $a_1^\dagger a_2$ and $a_2^\dagger 
a_1$. This gives the same susy $\vert 0 >$ that remains from 2 
D-branes at an angle as was discussed above. So it is consistent
with supersymmetry to turn on only this field.  

To show which modes become massless, and to display the localization
of states, we insert this background into the low energy
6-dimensional worldvolume Yang-Mills theory, and expand around
it. 
The important feature of the theory for this purpose is the 
d=6 potential $\tr [X^\mu,X^\nu]^2$
as well as the gauge couplings to the scalars
$\tr [A_\mu, X^\nu]$.  After rotation, we will derive an effective 4D
theory, in which the scalars and two of the 6D gauge fields become 
scalars
in 4D.
As in standard Kaluza-Klein, if we take the linear fluctuation to
be independent of the 4D
space-time, we obtain two-dimensional differential equations whose
eigenvalues are the masses of the 4D particles.
These 2D solutions are trapped inside a harmonic oscillator potential
well and are thus localized to the intersection of the branes. 

Let us briefly discuss a simple example,
the gauge field equation of motion.
The modes corresponding to open strings between the two branes 
are $\delta \phi=\Bigl(\matrix{0&f_1\cr 
f_2&0\cr}\Bigr)$ 
and $\delta A_{\mu}=\Bigl(\matrix{0&A^1_{\mu}\cr A^2_{\mu}&0\cr}\Bigr)$.
One finds, for example,
\eqn\hoeqn{
\partial^2\delta A_{\mu}^i+\left({\theta
\over \pi\alpha'}\right)^2{\vert z\vert}^2\delta A_{\mu}^i=0.
}
The background produces a space dependent mass $\theta | z|/\pi\alpha'$.
Clearly normal modes of \hoeqn\ will be localized around $z=0$;
they have the spacing for the tower of states found in section 3.

We used the $\alpha'\rightarrow 0$ limit on the 5-brane. How much
can we trust it? In
general, since there are linearly rising vev's, we expect that higher
derivative terms will be important; some of these would be taken into
account by considering the full Born-Infeld form of the action. 
Dangerous higher derivative terms are
those that, after taking one derivative of the vev's, are proportional
to the rotation angle and are constant throughout the $D=6$ spacetime,
and therefore not
suppressed by momentum power counting. Such terms
are independent of 6-dimensional space-time and so may shift the zero 
point
energy of the oscillator described above, but will not affect the
localization of the states. One might also expect that
a low energy effective action description can be trusted as
long as the states are not localized to within $\sqrt{\alpha'}$ of the 
intersection; hence, we assume $\theta<<1$.

\newsec{T-Duality}

\def\arg{{\rm Arg\hskip0.1em}}
\def\pn{\partial_n}
\def\pt{\partial_\tau}

We will now comment on T-duality.  
We consider our canonical example with two 2-branes, using the notation
of Section 2.2. The first plane lies along $\Re Z^1$, $\Re Z^2$,
while the second is rotated by $Z^1\to e^{i\theta} Z^1$,
$Z^2\to e^{-i\theta} Z^2$.

Open strings then have boundary conditions
\eqn\tdbctt
{\eqalign{
2_1:
&\;\;\;\;  \pn X^{6,8}=0\cr
&\;\;\;\;  \pt X^{7,9}=0\cr
2_2:
&\;\;\;\;  \pn\left( \cos\theta\; X^{6,8} \mp \sin\theta\; 
X^{7,9}\right)=0\cr
&\;\;\;\;  \pt\left( \pm\sin\theta\; X^{6,8} + \cos\theta\; X^{7,9}\right)=0.
}}
We consider toroidally compactifying $X^{6,7,8,9}$ on $T^4$, with
nontrivial metric components.
In particular, if we want the branes to be consistent with the toroidal
compactification, we should take the two $T^2$'s in $T^4$ to have
moduli $\arg\tau=\theta$.\foot{For simplicity, we consider here
a brane wrapped once around one cycle of the $T^2$.}

If we T-dual in the $X^{6,8}$ directions, we will get a square $T^4$
(K\"{a}hler moduli and complex structure being interchanged),
and the two $2$-branes will became a $0$-brane and a $4$-brane. The
boundary conditions become
\eqn\tdbctt
{\eqalign{
0:
&\;\;\;\;  \pt X^{6,8}=0\cr
&\;\;\;\;  \pt X^{7,9}=0\cr
4:
&\;\;\;\;  \pn X^{7,9}\mp\cot\theta\; \pt X^{6,8}=0\cr
&\;\;\;\;  \pn X^{6,8}\pm\cot\theta\; \pt X^{7,9}=0.
}}
As in Ref. \dbi, these boundary conditions can be thought of as a
$4$-brane in a background\foot{We define $\hat F=F+B$, where $F$ is
the worldvolume gauge field.}
\eqn\tdbg
{ \hat F_{76}=\hat F_{98}=-\cot\theta .}
Thus the supersymmetry condition that we found for the $2$-branes,
namely that the rotation angles for the two planes be equal, is
T-dual to an anti-self-dual gauge field background. To see that this
makes sense, consider the unbroken supersymmetries for the T-dual
system. Here we have:
\eqn\tdsusy
{\eqalign{
\tilde\eps=&\eps\cr
\tilde\eps=&-\Gamma^6 \Gamma^7 \Gamma^8 \Gamma^9 \eps.
}}
Defining the lowering operators $a_1=(\Gamma^6+i \Gamma^7)/2$ and
$a_2= (\Gamma^8+i \Gamma^9)/2$, the unbroken supersymmetries
are $\eps=\ket{0},a_1^\dagger a_2^\dagger \ket{0}$. These two
Fock states also satisfy $\Gamma^{67}\eps=+\Gamma^{89}\eps$.
Now in the presence of a background gauge field, the 
supersymmetry transformation of the gaugino is
\eqn\tdsusychi
{\delta\chi=\Gamma^{MN} \hat F_{MN}\;\eta}
Given the above remarks, this vanishes iff $ \hat F$ is anti-self-dual.
 
In other situations ($n>2$), the analogues of eqs. \tdsusy\
and \tdsusychi\ contain additional dependence on the background fields.
This will be explored in Ref. \vbrgl.

\newsec{Higher dimensional interpretation}

The two supersymmetries of the Type \IIa\ string are unified into a
single eleven-dimensional spinor $\eta$ in M-theory\refs{\mtheory,\duality}.
Does the condition \unb\ follow from something simple?

We now show that all of the D-brane supersymmetry conditions \unb, as 
well
as the condition for unbroken supersymmetry in the presence of a NS 
5-brane,
are unified into the condition
\eqn\munb{\eta = \prod_i e^\mu_i \Gamma_\mu\ \eta}
where the product over $e^\mu_i$ includes $e^{11}$ for a brane
with winding number around $X^{11}$ (such as the $4$-brane) 
or momentum $P^{11}$ (such as the $0$-brane),
but does not include $e^{11}$ for
others (such as the $2$-brane).

Proof: The condition \unb\ is
\eqn\eunb{
(1-\Gamma^{11})\eta = \Gamma_D (1+\Gamma^{11})\eta
}
where $\Gamma_D$ only includes dimensions within the ten.
Multiply both sides by $\Gamma_D$ and use
\eqn\square{
\Gamma_D^2 = \omega \equiv \cases{+1&for $k=4l+2$\cr -1&for $k=4l$}
}
to derive
\eqn\funb{
\omega(1+\Gamma^{11})\eta = \Gamma_D (1-\Gamma^{11})\eta.
}
Adding the two equations produces \munb, with $\Gamma^{11}$ appearing
in the product if $k=4l$.  

So, it clearly works for $D0$, $D2$,
$D4$ (wound $5$-brane) and the $9$-brane of Type \I.
The $d=11$ interpretation of $D8$ is still unclear \green.

The NS IIA 5-brane has chiral world brane susy. This means that the
two spinor projections are $\epsilon=\Gamma_D\epsilon$, $\tilde\epsilon
=\Gamma_D\tilde\epsilon$. Since $\eta=\epsilon+\tilde\epsilon$ then
$\eta=\Gamma_D\eta$ and indeed it is unwound. 

For the purposes of this criterion, the $D6$ brane is unwound.
As it is identified with the Kaluza-Klein monopole \town,
this may be a bit counterintuitive, but leads to no contradictions.

It would be quite interesting to propose a type \IIb\ version of
this.

\subsec{D-branes and discrete holonomy}

After compactification on a manifold $\CM$, the surviving supersymmetries
are those which are preserved by the holonomy group of $\CM$.
Can we generalize this statement to compactifications with branes?

A local test for unbroken susy is to scatter a low energy gravitino off 
of
the brane -- if it is unaffected, the associated susy is unbroken, but
if there is a phase shift, it is broken.
Since the phase shift is given by the action of the matrix $\Gamma_D$,
we want to regard this matrix as discrete holonomy.
As we saw above, in $d=11$ it is a linear operator on the spinor.

However, this will not be the holonomy of the metric-compatible 
connection,
because the condition for unbroken
supersymmetry also depends on the antisymmetric tensor background.
We know that we can have supersymmetric backgrounds with no covariantly 
constant spinor, but with zero modes of the full gravitino transformation
law.

The holonomy condition can be adapted to this situation by 
using the gravitino transformation law to define a modified connection,
as was proposed long ago for eleven-dimensional supergravity \duff.
The idea is that since the gravitino transformation law
\eqn\eltrans{
\delta \psi_M = \nabla_M \eps - {1\over 288} 
\left( {\Gamma_M}^{\,\,\, PQRS}-8 
{\delta}_M^P \Gamma^{QRS} \right) F_{PQRS} \ \eps .
}
is a linear operator on $\eps$, it can be regarded as a covariant
derivative written in terms of a modified connection $\Omega_M$.
In terms of the spin connection $\omega_M$ and $4$-form field strength 
$F$,
\eqn\modcon{
\Omega_M \equiv
{1\over 4}\omega_M^{mn}\Gamma_{mn} - {1\over 288} 
\left( {\Gamma_M}^{\,\,\, PQRS}-8 
{\delta}_M^P \Gamma^{QRS} \right) F_{PQRS} .
}
It defines the following parallel transport of a spinor:
\eqn\modpar{
0 = \dot X^M(t) \left( \nabla_M - {1\over 288} 
\left( {\Gamma_M}^{\,\,\, PQRS}-8 
{\delta}_M^P \Gamma^{QRS} \right) F_{PQRS}(X(t)) \right) \eps(X(t)) .
}
Integrating this along a path $X^M(t)$
defines the holonomy of the modified connection, or `generalized 
holonomy.'

We do not really have to go to $d=11$ for this -- we could
regard the two gravitino transformations in $d=10$ as defining a
connection which acts on the direct sum of the two spinor parameters.
This also
allows us to define generalized holonomy for the type \IIb\ theory.

In many cases the field strengths
can then be reduced to forms of rank $p\le 3$, and the modified connection
reduces to a connection with torsion, as in
\ref\stromtor{A. Strominger, ``{\it Superstrings with Torsion,}''
\npb{274}{1986}{253}.}.
However, this is not true in general.

To apply this to our D-brane case,
we should regard the condition \munb\ as defining a discrete 
generalized holonomy, concentrated at the position of the brane.  
The statement will then be that in general compactifications with branes,
unbroken supersymmetry must commute with the generalized holonomy group.

Does this agree with the generalized holonomy
of the solutions of the low energy
field theory as defined by \modpar ?
It is certainly clear from their BPS nature
that both generalized holonomy groups will preserve the same
supersymmetries.

We would expect the generalized holonomy group for the soliton
solutions to be larger, because the objects have a finite size.
Consider for example the supermembrane solution \duffstelle.
The $d=11$ supersymmetry parameter splits into two $d=8$ spinors,
and the gravitino transformation becomes (schematically)
$\delta\psi = (D+\Gamma_9 F)\eps$.
This cancels for one chirality, implying that the spin connection
and gauge field terms are equal, and thus the modified connection
acting on the other chirality spinor is equal to twice the spin connection.
Therefore the generalized holonomy group in this case is $SO(8)$.
It would be interesting to see this in the D-brane generalized holonomy 
group as well, by taking into account corrections in string
perturbation theory.

A priori the generalized holonomy group in a specific background
could be any subgroup of $SO(16,16)$, but one might expect
constraints on the possibilities.  This question is under investigation.

We believe generalized holonomy will be an important element in a
deeper understanding of the geometry of supergravity, as in
the works \nicolai.

\medskip

We are pleased to acknowledge stimulating conversations with A. Morosov,
I. Singer and A. Strominger.  M.R.D. thanks the LPTHE, Universit\'e de
Paris VI, for its hospitality.

\listrefs
\end